\documentclass[letter,twocolumn]{jpsj2} 

\title {Superconducting Gap and Pseudogap in Iron-Based Layered Superconductor La(O$_{1-x}$F$_x$)FeAs}

\author{Takafumi \textsc{Sato}$^{1}$\thanks{E-mail address: t-sato@arpes.phys.tohoku.ac.jp}, Seigo \textsc{Souma}$^{2,3}$, Kosuke \textsc{Nakayama}$^{1}$, Kensei \textsc{Terashima}$^{1}$, Katsuaki \textsc{Sugawara}$^{1}$, Takashi \textsc{Takahashi}$^{1,2,3}$, Yoichi \textsc{Kamihara}$^{4}$, Masahiro \textsc{Hirano}$^{4}$, and Hideo \textsc{Hosono}$^{4}$}

\inst{$^1$Department of Physics, Tohoku University, Sendai 980-8578\\
$^2$CREST, Japan Science and Technology Agency (JST), Kawaguchi 332-0012\\
$^3$WPI Research Center, Advanced Institute for Materials Research, Tohoku University, Sendai 980-8577\\
$^4$ERATO-SORST, Japan Science and Technology Agency (JST) and Frontier Research Center, Tokyo Institute of Technology, Nagatsuta 4259, Midori, Yokohama 226-8503}

\abst{We report high-resolution photoemission spectroscopy of newly-discovered iron-based layered superconductor La(O$_{0.93}$F$_{0.07}$)FeAs ($T_{\rm c}$ = 24 K).  We found that the superconducting gap shows a marked deviation from the isotropic $s$-wave symmetry.  The estimated gap size at 5 K is 3.6 meV in the $s$- or axial $p$-wave case, while it is 4.1 meV in the polar $p$- or $d$-wave case.  We also found a pseudogap of 15-20 meV above $T_{\rm c}$, which is gradually filled-in with increasing temperature and closes at temperature far above $T_{\rm c}$ similarly to copper-oxide high-temperature superconductors.}

\kword{superconducting gap, pseudogap, iron-based superconductor, photoemission spectroscopy}
\begin{document}
\maketitle

	Superconductivity observed at as high as 26 K (onset temperature = 32 K) in La(O$_{1-x}$F$_x$)FeAs \cite{ref1} provided a deep impact in condensed-matter physics, since this new superconductor does not belong to any known categories of ``high-temperature superconductors'' such as copper-oxides (cuprates)\cite{ref2}, fullerenes\cite{ref3}, and MgB$_2$\cite{ref4}.  La(O$_{1-x}$F$_x$)FeAs consists of alternatingly stacked insulating lanthanum oxide (LaO) and conductive iron arsenide (FeAs) layers.  While undoped LaOFeAs is a metal or a degenerate semiconductor at room temperature with no sign of superconductivity, substitution of oxygen with fluorine (F) atoms gives rise to the superconductivity at the F doping of $x$ = 0.03.  Further doping causes a gradual increase of superconducting transition temperature ($T_{\rm c}$) up to the maximum $T_{\rm c}$ of 26 K with the onset temperature over 30 K at $x$ = 0.11\cite{ref1}.  Although the crystal structure is substantially different from that of cuprate superconductors - it is not a perovskite structure and does not contain CuO$_2$ planes -, both compounds share intriguing similarities such as the two-dimensional crystal/electronic structure\cite{ref1,ref5,ref6,ref7,ref8}, presence of a superconducting dome in the electronic phase diagram where the $T_{\rm c}$ is controlled by a systematic aliovalent ion doping into the insulating block layers, and a characteristic anomaly in the transport property in the under-doped region\cite{ref1}.  It has been suggested that La(O$_{1-x}$F$_x$)FeAs is situated at the boundary to the ferromagnetic phase, since the F doping increases the number of Fe 3$d$ electrons from six to that (seven) of Co 3$d$ electrons in LaOCoAs which is known to undergo a ferromagnetic transition at 66 K.\cite{ref1,ref16}  Such a situation may invoke an exotic pairing mechanism related to ferromagnetic spin fluctuations.  It is thus particularly important to elucidate the mechanism and origin of superconductivity in this new category of high-$T_{\rm c}$ superconductor in relation to cuprates and other layered novel superconductors like ruthenium-\cite{ref9} and cobalt-oxides\cite{ref10}.

	In this letter, we report ultrahigh-resolution ($\Delta$E = 1.7 meV) photoemission spectroscopy (PES) on La(O$_{0.93}$F$_{0.07}$)FeAs to study the electronic structure and the superconducting gap.  We found that the PES spectrum at 5 K shows a suppression of spectral intensity within  4 meV indicative of opening of a superconducting gap.  We also observed a pseudogap with a finite density of states (DOS) at the Fermi level ($E_{\rm F}$), which opens at temperature even above $T_{\rm c}$ and closes at temperature far above $T_{\rm c}$.

	Polycrystalline La(O$_{0.93}$F$_{0.07}$)FeAs was synthesized by heating a mixture of lanthanum arsenide, iron arsenide, and dehydrated La$_2$O$_3$ powders in a silica tube filed with Ar gas at 1250$^{\circ}$C for 40 h.   Doping of F atoms was performed by adding 1:1 mixture of LaF$_3$ and La to the starting material.  The $T_{\rm c}$ (midpoint temperature = 24 K, onset temperature = 32 K) and the composition ($x$ = 0.07) were determined by the methods described in the previous work.\cite{ref1}  High-resolution PES measurements have been performed with a SES-2002 PES spectrometer with a high-intensity helium (He) discharge lamp equipped with a toroidal grating monochromator.  The He I$\alpha$ (21.218 eV) and He II$\alpha$ (40.814 eV) resonance lines were used to excite photoelectrons.  The energy resolution ($\Delta$E) was set at 1.7 meV except for the measurements of the wide valence-band region ($\Delta$E = 15 meV).  We fractured/scraped the sample under ultrahigh vacuum of 2$\times$10$^{-11}$Torr to obtain a clean and fresh surface for PES measurements.  We have confirmed that the degradation of sample surface did not take place during the measurements and the data shown here are reproducible by measuring several samples.  The Fermi level ($E_{\rm F}$) of sample was referenced to that of a gold film evaporated onto the sample holder.

\begin{figure}[t]
\begin{center}
\includegraphics[width=8cm]{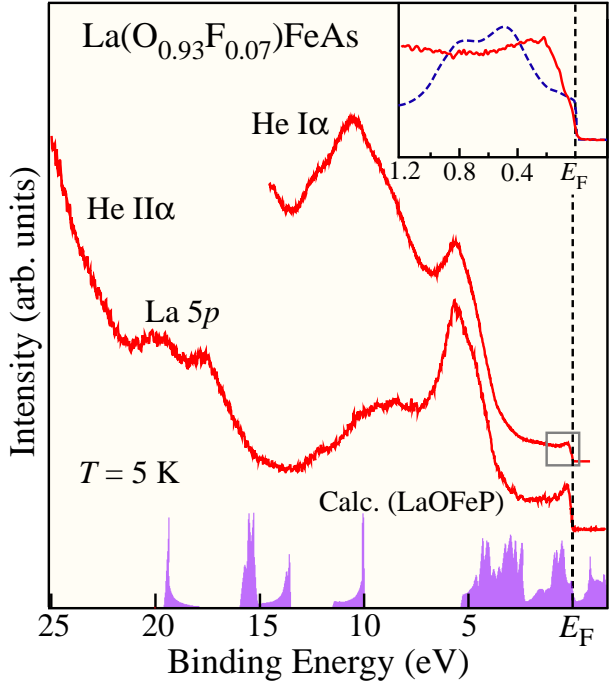}
\end{center}
\caption{Valence-band PES spectra of La(O$_{0.93}$F$_{0.07}$)FeAs measured at 5 K with He I$\alpha$ and He II$\alpha$ photons ($h\nu$ =21.218 and 40.814 eV).  The calculated DOS for LaOFeP (ref. 7) is also shown for comparison.  Inset shows an expansion near $E_{\rm F}$ of the He I spectrum, together with the calculated DOS (blue dashed line) broadened by a Gaussian with the full-width at half-maximum of 0.2 eV multiplied by the FD function.}
\label{f1}
\end{figure}

\begin{figure}[t]
\begin{center}
\includegraphics[width=8cm]{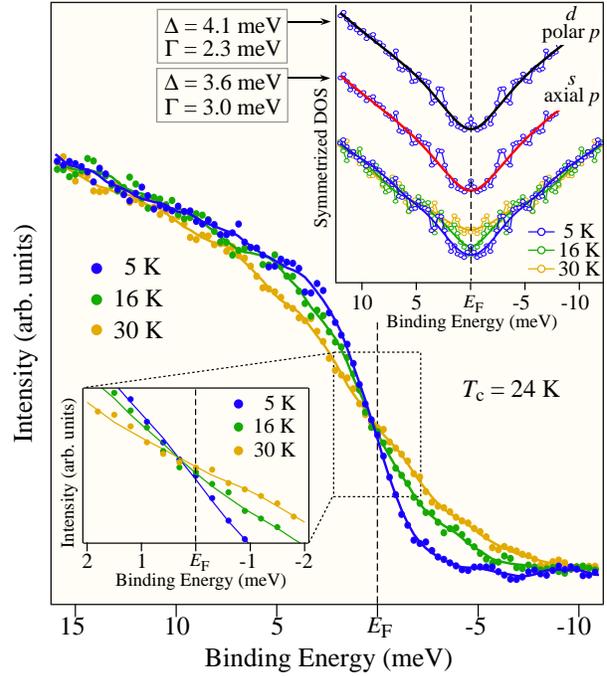}
\end{center}
\caption{High-resolution PES spectra near $E_{\rm F}$ of La(O$_{0.93}$F$_{0.07}$)FeAs measured at three temperatures across $T_{\rm c}$ with the He I$\alpha$ resonance line.  Expanded spectra in the vicinity of $E_{\rm F}$ are shown in the inset (left bottom).  Another inset (right top) shows the spectra symmetrized with respect to $E_{\rm F}$, together with the numerical fitting results to the symmetrized DOS at 5 K with the $s$-, $p$- (polar or axial), and $d$-wave gap functions (black and red curves).  We also measured PES spectra with the more bulk sensitive Xe I line (8.437 eV)\cite{ref13} and obtained essentially the same results.}
\label{f2}
\end{figure}

	Figure 1 shows valence-band PES spectra of La(O$_{0.93}$F$_{0.07}$)FeAs (slightly under-doped, $T_{\rm c}$ = 24 K) measured at 5 K with the He I$\alpha$ and He II$\alpha$ resonance lines.  We find several peaks in both spectra, such as a prominent peak at 5.5 eV accompanied by a shoulder structure at lower binding energy, a broad feature at 7-13 eV composed of multiple peaks, and a small peak in the vicinity of $E_{\rm F}$, while the intensity of these peaks is considerably different between the He I and He II spectra.  The He II spectrum shows an additional large doublet structure at around 19 eV, which is assigned to the spin-split La 5$p$ core level.\cite{ref11}  We also display in Fig. 1 the calculated DOS for LaOFeP using the density functional theory.\cite{ref7}  Since the band calculation of LaOFeAs is not available at present, we compare the present PES result with the calculation for LaOFeP by taking account of the possible difference between the 3$p$ (P) and 4$p$ (As) states.  The calculated electronic states near $E_{\rm F}$ are dominated by the Fe 3$d$ states, while the states at 2-5 eV are mainly from the O 2$p$ states with admixture from the Fe 3$d$ and P 3$p$ states.  We therefore attribute the near-$E_{\rm F}$ states and the 5.5-eV feature in the PES spectrum to the Fe-3$d$ and O-2$p$ dominated states, respectively.  This assignment is supported by the experimental fact that the intensity of the near-$E_{\rm F}$ peak is enhanced with respect to the 5.5-eV peak in the He II spectrum, since the ratio of photoionization cross-section between the Fe 3$d$ and O 2$p$ orbital for the He II photons is about twice as large as that for the He I photons.\cite{ref12}  The peaks at 7-13 eV in the spectrum may be attributed to the As 4$s$ states since the calculated P 3$s$ states are also in this energy range (10-12 eV).  The finite difference in the energy position of O-2$p$ dominated states between the PES spectrum and the calculated DOS may be due to underestimation of the binding energy of O 2$p$ states in the LDA (local-density-approximation) calculation.  On the other hand, the location of Fe 3$d$ states near $E_{\rm F}$ shows a good agreement between the experiment and the calculation.  It is noted that the peak position of near-$E_{\rm F}$ states is not at $E_{\rm F}$ but at 0.2 eV away from $E_{\rm F}$, as clearly visible in the expanded He I spectrum near $E_{\rm F}$ (inset to Fig. 1), demonstrating that the DOS gradually decreases with approaching $E_{\rm F}$.  This feature is also seen in the calculation although the slope of DOS toward $E_{\rm F}$ is steeper in the experiment.

	Figure 2 shows ultrahigh-resolution PES spectra near $E_{\rm F}$ of La(O$_{0.93}$F$_{0.07}$)FeAs measured at three temperatures across $T_{\rm c}$ with the He I$\alpha$ resonance line.  We find that the crossing point of PES spectra  is not at $E_{\rm F}$ but at about 0.4 meV higher binding energy with respect to $E_{\rm F}$ as clearly seen in the inset (left bottom), where the PES spectra in the vicinity of $E_{\rm F}$ are expanded.  Further, the spectral intensity at $E_{\rm F}$ gradually increases with increasing temperature.  These results indicate the opening of a superconducting gap.  We do not observe a sharp superconducting coherence peak, which suggests that the gap function has a marked deviation from the conventional isotropic $s$-wave symmetry.  We also find a considerable spectral intensity at $E_{\rm F}$ even well below $T_{\rm c}$, which suggests (i) multi-band superconductivity where some parts of Fermi surfaces remain ungapped even in the superconducting state, (ii) a metallic non-superconducting portion of the sample, and/or (iii) presence of line/point nodes as in the $p$- and $d$-wave cases.  We symmetrized the PES spectra with respect to $E_{\rm F}$ to see more clearly the superconducting gap.  The result is shown in the inset (right top) to Fig. 2.  The symmetrized PES spectra at 5 K and 16 K show a suppression of intensity within $\sim$4 meV centered at $E_{\rm F}$ compared to that at 30 K, suggesting the opening of a superconducting gap.  We numerically fit the symmetrized spectrum at 5 K to estimate the superconducting-gap size with the $s$-, $p$-, or $d$-wave gap function by assuming the two-dimensional cylindrical Fermi surface.  It is known that the shape of DOS for the axial (polar) $p$ wave\cite{ref14} is the same as that of $s$- ($d$-) wave for the case of cylindrical Fermi surface.  We fit the spectrum by the Dynes function\cite{ref15} which is convoluted with a Gaussian to take into account the instrumental resolution.  As seen in the inset to Fig. 2, both fitting curves with different symmetries look to reasonably reproduce the experimental spectrum.  In the $s$- and axial $p$-wave cases, the gap size $\Delta$ is estimated to be 3.6$\pm$0.3 meV with the broadening factor $\Gamma$ of 3.0 meV, while in the polor $p$- and $d$-wave cases, the $\Delta$ is 4.1$\pm$0.3 meV with the $\Gamma$ value of 2.3 meV.  It is remarked that, in the $s$- and axial $p$-wave cases, the $\Gamma$ value is relatively large and even comparable to the gap size.  This again indicates that the superconducting gap of La(O$_{0.93}$F$_{0.07}$)FeAs is substantially deviated from the isotropic $s$-wave symmetry, suggesting higher possibility for the polar $p$- or $d$-wave symmetry.  In this case, the reduced gap value 2$\Delta$(0)/$k_{\rm B}$$T_{\rm c}$ estimated with the obtained gap value is about 4.
	
\begin{figure}[t]
\begin{center}
\includegraphics[width=8cm]{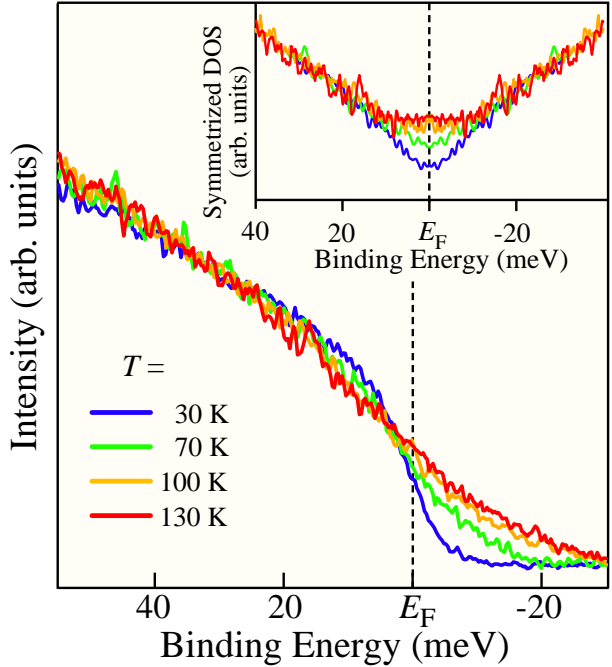}
\end{center}
\caption{High-resolution PES spectra near $E_{\rm F}$ measured at several temperatures above $T_{\rm c}$ (30-130 K) with He I$\alpha$ line.  Inset shows the spectra symmetrized with respect to $E_{\rm F}$.}
\label{f3}
\end{figure}

	Figure 3 displays temperature dependence of ultrahigh-resolution PES spectrum near $E_{\rm F}$ of La(O$_{0.93}$F$_{0.07}$)FeAs measured at several temperatures above $T_{\rm c}$ (30-130 K).  Surprisingly, the spectral intensity at $E_{\rm F}$ increases with increasing temperature even far above $T_{\rm c}$, unlike the temperature dependence of the FD function where the intensity at $E_{\rm F}$ always keeps a constant value.  We have confirmed this spectral change being intrinsic and not due to the extrinsic effect such as sample degradation, by cycling the temperature of measurements.  This unambiguously indicates that the temperature induced spectral change of La(O$_{0.93}$F$_{0.07}$)FeAs above $T_{\rm c}$ is due not only to the FD function but also to a systematic evolution of the spectral weight in the vicinity of $E_{\rm F}$.  To see more directly the temperature-induced change of DOS near $E_{\rm F}$, we symmetrized each PES spectrum to remove the effect of FD function, and show the result in the inset.  As seen in the figure, the spectral DOS at the energy higher than 15-20 meV is almost independent of temperature.  In contrast, the DOS near $E_{\rm F}$ is remarkably suppressed up to about 15-20 meV binding energy, suggesting that a pseudogap with a finite DOS at $E_{\rm F}$ opens above $T_{\rm c}$ and is gradually filled-in with increasing temperature.  It is noted that this pseudogap is different from the superconducting gap because the energy scale is apparently different from each other.  A similar behavior in the temperature dependence of DOS has been also observed in La$_{1.85}$Sr$_{0.15}$CuO$_4$,\cite{satoLSCO} which exhibits the superconducting gap of 8 meV and the pseudogap of 30-35 meV.  The symmetrized PES spectrum at $T$ = 100 K almost coincides with that at 130 K, suggesting that the pseudogap closes at around $T$* = 100 K.  It is remarked that this $T$* value is similar to the temperature where the electrical resistivity shows a characteristic reduction in underdoped region,\cite{ref1} suggesting that the anomaly in the resistivity is closely related to the opening of pseudogap.
	
	Although the origin of pseudogap is not clear at present, several candidates are considered.  One plausible candidate is a ferro- or antiferro-magnetic spin correlation.  Another scenario to explain the pseudogap is the precursor pairing, where the electron pairs without phase coherence are expected to exist above $T_{\rm c}$.  In this case, the observed $T$* of 100 K suggests a potentially higher $T_{\rm c}$ to be achieved in La(O$_{1-x}$F$_x$)FeAs.  Present PES results provide an important experimental basis to construct a microscopic theory to explain the mechanism of superconductivity in this iron-based superconductor.
	

     In conclusion, we reported high-resolution PES study of newly-discovered iron-based superconductor La(O$_{0.93}$F$_{0.07}$)FeAs.  The superconducting gap shows a marked deviation from the isotropic $s$-wave symmetry, and the gap size is estimated to be $\sim$4 meV at 5 K.  We also found that a pseudogap of 15-20 meV opens at $E_{\rm F}$ even above $T_{\rm c}$ and closes at temperature far above $T_{\rm c}$, similarly to cuprate high-temperature superconductors.

\section*{Acknowledgments}
We thank K. Ishida for useful discussions.  We also thank T. Arakane, Y. Sekiba, and K. Mizuno for their help in the experiment.  This work was supported by grants from JSPS, JST-CREST, and MEXT of Japan.

\end{document}